\documentclass{article}

\usepackage{arxiv}

\usepackage[utf8]{inputenc} 
\usepackage[T1]{fontenc}    
\usepackage{hyperref}       
\usepackage{url}            
\usepackage{booktabs}       
\usepackage{amsfonts}       
\usepackage{nicefrac}       
\usepackage{microtype}      
\usepackage{lipsum}		
\usepackage{microtype,mathtools}
\usepackage{amsmath,amssymb,amsthm}
\usepackage{graphicx}
\usepackage{natbib}
\title{Turbulent characteristics within roughness sublayer over spanwise homogeneous transversal bars}

\author{
  Chao Wang, Ph.D., \\
  The University of Texas at Dallas\\
  Richardson, TX\\
  \texttt{cxw151530@utdallas.edu} \\
    \And
  Ching Cheng, P.E., \\
  Jacobs Engineering Group\\
  Dallas, TX\\
  \texttt{Ching.Cheng@jacobs.com} \\
}

\begin{document}
\maketitle

\begin{abstract}
Turbulent characteristics within shear layer have been studied, recently, within vegetative canopy, buildings, dunes. Kevin-Helmholtz instability triggered hairpin vortex shedding has been widely concluded as the ``signature'' of mixing layer analogy. However, convoluted roughness types complicate the observing practice of turbulent evolving progress. To simplify that, spanwise homogeneous roughness has been adopted in this work to capture the streamwise flow alternations under different streamwise roughness element distances. Streamwise successive object distance $\lambda_x$ is used to show the roughness element distance. The transitional trends from turbulent channel flow to canopy flow have been observed from Case 1 to Case 4. However, when $\lambda_x/h<1.0$, as successive distance decreases, the evidences indicate turbulent flow is transferring from canopy flow into channel flow from Case 4 to Case 6. Instantaneous results verifies the enhancement of turbulent mixing and decreasing turbulent coherency. The attached-eddy hypothesis (AEH) becomes valid in $\lambda_x/h<1.0$ Cases. Meanwhile, Reynolds-averaged flow results show the roughness shear ``curve'' effectiveness will decrease when $\lambda_x$ keeps decreasing after $\lambda_x/h<1.0$. The probability density function (PDF) of three fluctuating variables on a sampling point over element display the ``increasing-decreasing'' trend of decreasing spacing effect on upper region, wherein the maximal effect emerges in Case 2 and 3. And the swirling structural visualizations also demonstrate the progress of turbulent structure changing in roughness sublayer at different streamwise successive object distances. All results comprehensively show a great consistent evidence, that is the channel flow-analogy emerges dense roughness canopy when $\lambda_x<1.0$, where topographies show roughness effects. Finally, the instantaneous results show wall-normal momentum flux transports and turbulence swirling structures are correlated to the element spacing. 
\end{abstract}


\section{Introduction}
The prevalent studies of turbulent boundary layer over different types of roughness elements (building, vegetation, dunes and so forth) have highlighted the significance of turbulent characteristics in geophysics and engineering. Recently, the roughness sublayer (RSL) attributes in various canopy flows have been well studied \citep{bailey2016creation,wang2019turbulence,pan2014large,CocealBelcher04,Macdonald00,yangetal16,baietal15,bohm13,wang2018large}. All of these work have indicated the roughness sublayer depth $z_{RSL}$ can ranges three to five roughness element heights $h$, $2\lesssim z_{RSL}/h\lesssim5$, wherein turbulent mixing process is much more vigorous than the outer layer flow \citep{AndersonMeneveau11,AndersonMeneveau2010,wang2019turbulence,AndersonChamecki14}. The obstructive effects of canopy elements induce strong shear layer showing mixing-layer turbulent characteristics such as vorticity thickness scaled turbulent coherence length scales and hairpin vortex shedding which has been widely recorded \citep{bailey2016creation,AndersonChamecki14,wang2016numerical,wang2017large,wang2017numerical,bristow2017experimental,wang2018large,wang2019turbulence,wang2019numerical2,wang2018large2,wang2019thesis}. Inertial sublayer overlaid on the roughness sublayer maintained the channel-flow properties such as Attached-Eddy Hypothesis (AEH), where the integral length scale linearly increases with wall-normal elevation, and ineffectiveness of bottom topographies has been validated in idealized and realistic cases by \citep{wang2019turbulence}. Given the complicacy of natural geological digital elevation map, the idealized case is adopted as an isolation from such complicacy. Although consistency is not diluted by the obviously different element distributions,  the effects of topography distribution on boundary layer structure and turbulent coherency have not been fully demonstrated. Realistic simulation shows a series of ``field-like'' statistics, wherein the roughness elements effect to the aloft boundary layer structure is more evident. In this work, the roughness element spatial distribution effects on RSL will be discussed via a series of spanwise homogeneous roughness elements. 

Canopy successive spacing effects, previously, have been discussed \citep{bailey2013turbulence,wang2019thesis,hongetal11,AndersonChamecki14,bohm13,leonardietal03}. \citep{bailey2013turbulence} has displayed the transitional turbulent flow behaviors in vegetative canopy. According to the heterogeneous canopy simulation, significant dispersive fluxes have been oberseved to arise as a result of plant-scale heterogeneity, which have been proved to generate a strong influence on the mean velocity field, particularly the vertical velocity. In Figure 5, the instantanous velocity results also confirm such strong influence on the wall-normal velocity heterogeneous pattern. \citep{bailey2013turbulence} captures the flow behaved predominantly as a canopy layer although the integral lengths lose the mixing-layer analogy in the sparse canopies, where the mixing-layer analogy has been widely approved in vegetative, dune canopies \citep{bailey2013turbulence,wang2019thesis,wang2019turbulence,bailey2016creation}. \citep{wang2019thesis} demonstrated mixing layer analogy within dune field canopy, where the wall-normal heterogeneity is triggered by the local dune obstruction. And \citep{wang2019turbulence} calculated integral length profiles in dune field with different proximal distances. The results proved the strong correlation between turbulent coherency and $\lambda_x$. Leeward shear flow will grow in length scale in the downwelling advection. Downwelling shear eddies shows attached-eddy property, which is the length scale grows with local wall-normal elevation. Spatially correlated turbulent eddies can impinge on the downwind surfaces and break into eddies with small length scales. However, in these pre-existing works, turbulent swirling structure has not been displayed associated with turbulent evolving progresses, under monotonically modified successive element spacings. \citep{AndersonChamecki14} observed the streamwise turbulent swirling structure is the signatures of Kelvin-Helmholtz instabilities, which are responsible for enhanced downward transport of high-momentum fluid via the turbulent sweep mechanism. Meanwhile, the spatial structure of Kelvin-Helmholtz instabilities present the streamwise structure are mixing-layer vortex rollers. 

Various simplified roughness canopies have been adopted to study streamwise heterogeneity within shear induced turbulent mixing progress. \citep{leonardietal03} and \citep{leonardi2004structure} have recorded the DNS results over transversal bars, showing a very effective consequence to isolate spanwise inhomogeneity. Direct Numerical Simulation (DNS), veraciously, recovers wall-normal instabilities and autocorrelation magnitudes between each roughness element. The pressure and viscous drag ratio, $w/k$ has been found to be highly correlated with element streamwise distance \citep{leonardietal03}, where $w$ is the streamwise distance between successive transversal bar and $k$ is the roughness height. As the same isolation in this work, transversal bars are also used to suppress the spanwise induced inhomogeneity and instability. \citep{leonardi2004structure} reports that as $w/k$ increases, the coherence decreases in the streamwise direction having a minimum for $w/k=7$. That is because most outward motion occurring close to the leading edge of the elements. Meanwhile, the Reynolds stress anisotropy tensor and its invariants show a closer approach to isotropy over the rough wall than over a smooth wall. In this work, the LES results will display the coherence decreasing trend in the streamwise direction as the spatial spacing decreases. Meanwhile, the three dimensional swirling structures reveal such coherences are evolving to smaller coherent spanwise vortices with positive rotating direction. 

In this work, a series of topographies, composed of spanwise transversal bars, are used as Case 1 to Case 6. Streamwise spacing $\lambda_x$ indicate the successive streamwise distance between each transversal bar, wherein $\lambda_x$ is constant in each case (refer Figure 1). Section \ref{s:les_case} introduce the general backward of Large-Eddy Simulation method and the details of different cases. While, the results are displayed in Section \ref{s:results}. Statistical and instantaneous velocity field are displayed in Section \ref{s:results}. The turbulent coherences are well captured in the Reynolds-averaged $Q$ criterion. The PDF of streamwise and wall-normal velocity fluctuation consistently indicate the characteristics of roughness sublayer changes. The wall stress fluctuation PDF verifies the alternations of flow attributes between channel flow and canopy flow properties with $\lambda_x$ decreasing. Section \ref{s:conclusion} is displayed as a enclosure of this work at the end. 

\section{Large-Eddy Simulation \& Cases}
\label{s:les_case}

Recently, Computational Fluid Dynamics (CFD) method benefits a wide range of researches such as biomedicine, environment, geoscience and so forth, wherein different length scales of fluid are involved in these CFD simulations, especially in turbulent flow which consisted of a broad range of spectrum \citep{wang2019thesis}. To recover the geophysical fluid field, Direct Numerical Simulation (DNS) could be a promising methodology, resolving turbulence from Kolmogorov scale to the maximum of numerical length scale \citep{pope00}. However, DNS is too ``expensive'' for most geophysical fluid simulations due to the critical computational resource need in high Reynolds number flow regime. In contrast, LES can approach highly credible results with an acceptable numerical requirement \citep{metias96}. In this chapter, Large-Eddy Simulation method will be detailed discussed. Meanwhile, in Section \ref{sb:cases}, all numerical cases details will be provided. 

\subsection{Large-Eddy Simulation}
\label{sb:les}
In Large-Eddy Simulation (LES) method, the filtered three-dimensional transport equation, incompressible momentum, 
\begin{equation}
D_t \tilde{\boldsymbol u} ({\boldsymbol x},t) = \rho^{-1} {\boldsymbol F}({\boldsymbol x},t), 
\end{equation}
is solved, where $\rho$ is density, $\tilde{.}$ denotes a grid-filtered quantity, $\boldsymbol{u}(\boldsymbol{x},t)$ is velocity (in this work, $u$, $v$, $w$ are corresponded to velocity in streamwise, spanwise and wall-normal direction, respectively) and $\boldsymbol{F}(\boldsymbol{x},t)$ is the collection of forces (pressure correction, pressure gradient, stress heterogeneity and obstacle forces). The grid-filtering operation is attained here via convolution with the spatial filtering kernel, $\tilde{\boldsymbol u} ({\boldsymbol x},t) = G_\Delta \star {\boldsymbol u} ({\boldsymbol x},t)$, or in the following form
\begin{equation}
\tilde{\boldsymbol{u}}(\boldsymbol{x},t)=\oint G_{\Delta}(\boldsymbol{x}-\boldsymbol{x}^{\prime},t)\boldsymbol{u}(\boldsymbol{x}^{\prime},t)\mathrm{d}\boldsymbol{x}^{\prime},
\end{equation}
where $\Delta$ is the filter scale \citep{meneveaukatz}. A right-hand side forcing term, $- \nabla \cdot \boldsymbol{\mathsf{T}}$, will be generated after the filtering operation to momentum equation, where $\boldsymbol{\mathsf{T}} = \langle {\boldsymbol u}^\prime \otimes {\boldsymbol u}^\prime \rangle_t$ is the subgrid-scale stress tensor and $\langle . \rangle_a$ denotes averaging over dimension, $a$ (in this article, rank-1 and -2 tensors are denoted with bold-italic and bold-sans relief, respectively).

For the present study, $D_t \tilde{\boldsymbol u} ({\boldsymbol x},t) = \rho^{-1} {\boldsymbol F} ({\boldsymbol x},t)$ is solved for a channel-flow arrangement \citep{albertsonparlange1999,AndersonChamecki14}, with the flow forced by a pressure gradient in streamwise direction, $\boldsymbol{\mathrm{\Pi}} = \{\mathrm{\Pi},0,0\}$, where 
\begin{equation}
\mathrm{\Pi} = \left[ \mathrm{d}P_{0}/\mathrm{d}x\right]\frac{H}{\rho}=\tau^{w}/\rho=u_{*}^{2}=1,
\end{equation}
which sets the shear velocity, $u_{*}$, upon which all velocities are non-dimensionalized. In simulation, all length scales are normalized by $H$, which is the surface layer depth, and velocity are normalized by surface shear velocity. $D_t \tilde{\boldsymbol u} ({\boldsymbol x},t) = \rho^{-1} {\boldsymbol F} ({\boldsymbol x},t)$ is solved for high-Reynolds number, fully-rough conditions \citep{Jimenez2004}, and thus viscous effects can be neglected in simulation, $\nu \nabla^2 \tilde{\boldsymbol u} ({\boldsymbol x},t) = 0$. Under the presumption of $\rho ({\boldsymbol x},t) \rightarrow \rho$, the velocity vector is solenoidal, $\nabla \cdot \tilde{\boldsymbol u} ({\boldsymbol x},t) = 0$. During LES, the (dynamic) pressure needed to preserve $\nabla \cdot \tilde{\boldsymbol u} ({\boldsymbol x},t) = 0$ is dynamically computed by computation of $\nabla \cdot \left[ D_t \tilde{\boldsymbol u} ({\boldsymbol x},t) = \rho^{-1} {\boldsymbol F} ({\boldsymbol x},t) \right]$ and imposing $\nabla \cdot \tilde{\boldsymbol u} ({\boldsymbol x},t) = 0$, which yields a resultant pressure Poisson equation. 

The channel-flow configuration is created by the aforementioned pressure-gradient forcing, and the following boundary condition prescription: at the domain top, the zero-stress Neumann boundary condition is imposed on streamwise and spanwise velocity, $\partial \tilde{u} / \partial z | _{z / H = 1} = \partial \tilde{v} / \partial z | _{z / H = 1} = 0$. The zero vertical velocity condition is imposed at the domain top and bottom, $\tilde{w} (x , y , z / H = 0) = \tilde{w} (x,y,z / H = 1) = 0$. Spectral discretization is used in the horizontal directions, thus imposing periodic boundary conditions on the vertical ``faces'' of the domain, \textit{vis.}
\begin{equation}
\phi(x+mL_x,y+nL_y,z)=\phi(x,y,z), 
\end{equation}
and imposing spatial homogeneity in the horizontal dimensions. The code uses a staggered-grid formulation \citep{albertsonparlange1999}, where the first grid points for $\tilde{u}({\boldsymbol x},t)$ and $\tilde{v}({\boldsymbol x},t)$ are located at $\delta z/2$, where $\delta z = H/N_z$ is the resolution of the computational mesh in the vertical ($N_z$ is the number of vertical grid points). Grid resolution in the streamwise and spanwise direction is $\delta x = L_x/N_x$ and $\delta y = L_y/N_y$, respectively, where $L$ and $N$ denote horizontal domain extent and corresponding number of grid points (subscript $x$ or $y$ denotes streamwise or spanwise direction, respectively). Table 1 provides a summary of the domain attributes for the different cases, where the domain height has been set to the depth of the surface layer, $L_z/H = 1$.

At the lower boundary, surface momentum fluxes are prescribed with a hybrid scheme leveraging an immersed-boundary method (IBM)\citep{AndersonMeneveau2010,Anderson12} and the equilibrium logarithmic model \citep{piomelli} , depending on the digital elevation map, $h(x,y)$. When $h(x,y) < \delta z/2$, the topography vertically unresolved, and the logarithmic law is used:
\begin{equation} \label{Equation1}
\tau_{xz}^{w}(x,y,t)=-\left[{\frac{{\kappa}U(x,y,t)}{\log(\frac{1}{2}\delta{z}/\hat{z}_{0})}}\right]^{2}\frac{\bar{\tilde{u}}(x,y,\frac{1}{2}\delta{z},t)}{U(x,y,t)}
\end{equation}
and
\begin{equation} \label{Equation2}
\tau_{yz}^{w}(x,y,t)=-\left[{\frac{{\kappa}U(x,y,t)}{\log(\frac{1}{2}\delta{z}/\hat{z}_{0})}}\right]^{2}\frac{\bar{\tilde{v}}(x,y,\frac{1}{2}\delta{z},t)}{U(x,y,t)}
\end{equation}
where $\hat{z}_{0}/H=2\times10^{-4}$ is a prescribed roughness length, $\bar{\tilde{.}}$ denotes test-filtering \citep{germanoJFM,germanoetal} (used here to attenuate un-physical local surface stress fluctuations associated with localized application of Equation \ref{Equation1} and \ref{Equation2} \citep{bouzeidetal2005a}), and $U(x,y,\frac{1}{2}\delta{z},t)=(\bar{\tilde{u}}(x,y,\frac{1}{2}\delta{z},t)^{2}+\bar{\tilde{v}}(x,y,\frac{1}{2}\delta{z},t)^{2})^{1/2}$ is magnitude of the test-filtered velocity vector. Where $h(x,y) > \frac{1}{2}\delta{z}$, a continuous forcing Iboldsymbol is used \citep{Anderson12,mittaliaccarino2005}, which has been successfully used in similar studies of turbulent obstructed shear flows \citep{AndersonChamecki14,Andersonetal15b,anderson16}. The immersed boundary method computes a body force, which imposes circumferential momentum fluxes at computational ``cut'' cells based on spatial gradients of $h(x,y)$:
\begin{equation}
{\boldsymbol f}({\boldsymbol x},t)=-\frac{\tilde{\boldsymbol{u}}(\boldsymbol{x},t)}{\delta z} R(\tilde{\boldsymbol{u}}(\boldsymbol{x},t)\cdot \nabla h), 
\end{equation}
where $R$ is called Ramp Function \citep{Anderson12}
\begin{equation}
R(x)=\begin{cases}
    x & \text{for $x>0$},\\
    0 & \text{for $x\leqslant0$}.
  \end{cases}
\end{equation}
 Equations \ref{Equation1} and \ref{Equation2} are needed to ensure surface stress is imposed when $h(x,y) < \frac{1}{2}\delta{z}$. Subgrid-scale stresses are modeled with an eddy-viscosity model, 
\begin{equation}
{\boldsymbol \tau}^{d}=-2\nu_{t}\boldsymbol{\mathsf{S}}, 
\end{equation}
where 
\begin{equation}
\boldsymbol{\mathsf{S}}=\frac{1}{2}(\nabla {\tilde{\boldsymbol u}}+\nabla {\tilde{\boldsymbol u}}^\mathrm{T}) 
\end{equation}
is the resolved strain-rate tensor. The eddy viscosity is 
\begin{equation}
\nu_{t}=(C_{s}\Delta)^{2}|\boldsymbol{\mathsf{S}}|, 
\end{equation}
where $|\boldsymbol{\mathsf{S}}|=(2\boldsymbol{\mathsf{S}} \boldsymbol{:} \boldsymbol{\mathsf{S}})^{1/2}$, $C_{s}$ is the Smagorinsky coefficient, and $\Delta$ is the grid resolution. For the present simulations, the Lagrangian scale-dependent dynamic model is used \citep{bouzeidetal2005a}. The simulations have been run for $N_t \delta_t U_0 u_{*,d} H^{-1} \approx 10^3$ large-eddy turnovers, where $U_{0}=\langle{\tilde{u}(x,y,(L_z - \delta z/2)/H=1,t)}\rangle_{t}$ is a ``free stream'' or centerline velocity. This duration is sufficient for computation of Reynolds-averaged quantities.

\subsection{Cases}
\label{sb:cases}

To retrieve the streamwise heterogeneity, a series of numerical cases are used in this work. Figure 1 shows the numerical spatial information. All cases share the same size of domain, $L_x=L_y=2\pi H$ and $L_z=H$, where $L_x$, $L_y$ and $L_z$ are streamwise, spanwise and wall-normal spatial extent, $H$ is domain height, $H=50 m$. The mesh grid is fine enough to capture the major turbulent coherent structures and fluctuations, $\Delta_x=\Delta_y=0.05H=2.45m$ and $\Delta_z=0.008H=0.39m$. The large-eddy turnovers $N_t\delta_tU_0u_{*,d}H^{-1}\gtrsim 2\times10^{3}$ for each cases. This duration is sufficient for computation of Reynolds-averaged quantities. A streamwise successive object distance $\lambda_x$ is leveraged to control the roughness element density. From Case 1 to 6, $\lambda_x/H=2.94, 2.16, 1.37, 0.98, 0.59, 0.20$, respectively. 

Figure 1 shows the numerical case details. The black region indicates the area with zero elevation. The white zones are transversal bars with $h/H=0.1$, or $h=5m$. $\lambda_x$ is showed in Panel (a), which stars from the leeward face of first element to the leeward face of second element. Sparse elements are distributed in Panel (a) to (c). While, with $\lambda_x$ decreasing, the streamwise spacing is close to the transversal bar width in Panel (f). In Section \ref{s:results}, we can see the Panel (f) starts to show channel flow turbulent attributes. All numerical information is summarized in Table 1. Table 2 is displayed here to show all variables used in the work for the readers' convenience. 

\begin{figure}
\begin{center}
\includegraphics[width=16.5cm]{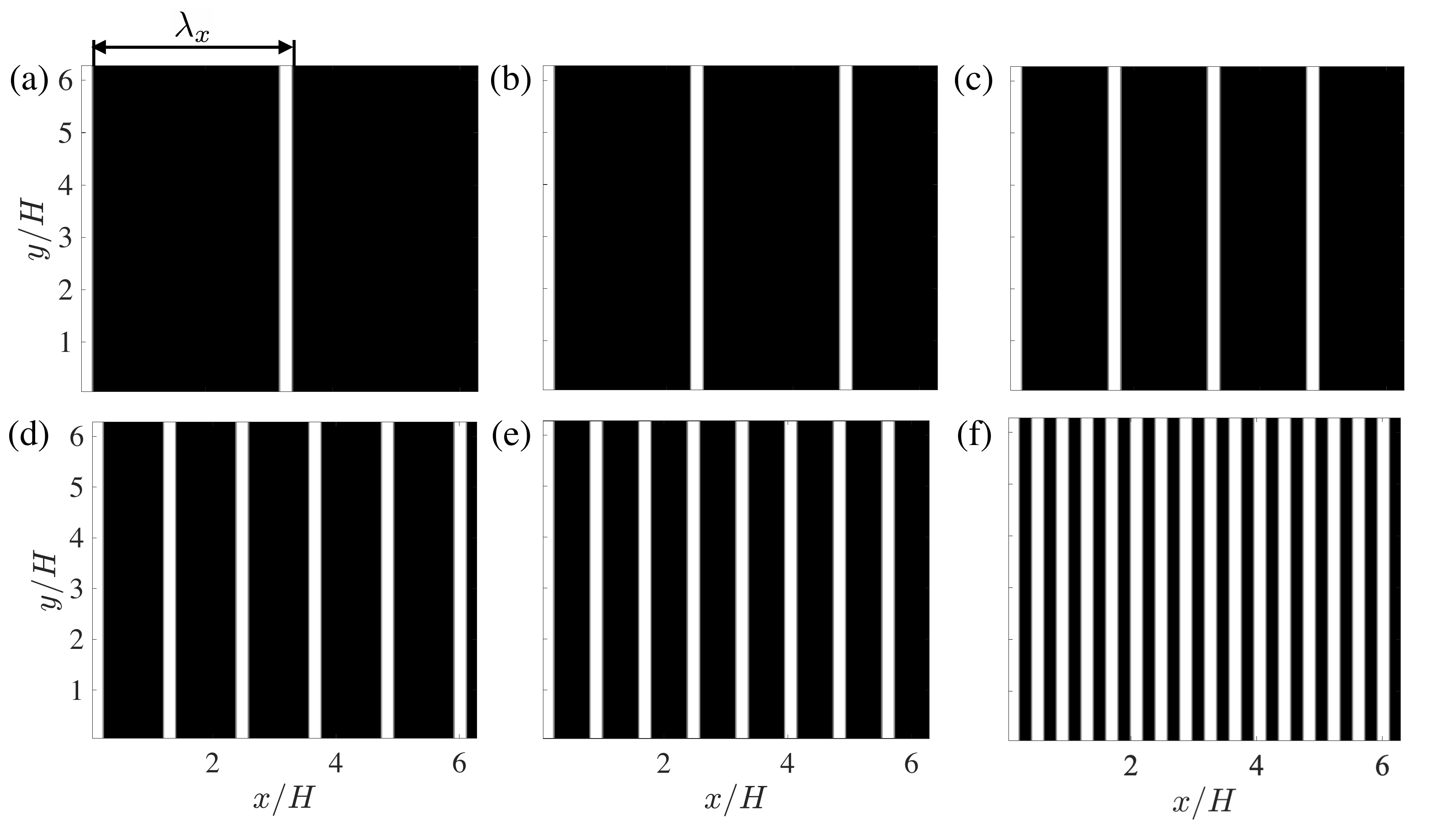}
\caption{Numerical cases used in this work. Panel (a) to (f) are Case 1 to Case 6 respectively. White color denotes the spanwise extending bars which is used to main spanwise homogeneity. The height of the object is $h=0.1H=5m$. While streamwise heterogeneity is triggered by various streamwise successive spacing $\lambda_x$. From Case 1 to 6, $\lambda_x/H=2.94, 2.16, 1.37, 0.98, 0.59, 0.20$. respectively.}
\label{Figure1}
\end{center}
\end{figure}

\begin{table*} 
\caption{Summary of numerical simulation and different cases attributes in this work ($H$ is domain height, $H=50 m$).}
\begin{center}
\begin{tabular}{cccccccccc}
\hline
Case 	& $L_x/H$		& $L_y/H$		& $L_z/H$		& $N_x$		& $N_y$		& $N_z$		& $h/H$	& $\lambda_x/H$ 	& $z_0/H$\\
Case 1	& $2\pi$		& $2\pi$		& $1.00$		& $128$		& $128$		& $128$		& $0.1$	& $2.94$			& $2\times10^{-4}$\\
Case 2	& $2\pi$		& $2\pi$		& $1.00$		& $128$		& $128$		& $128$		& $0.1$	& $2.16$			& $2\times10^{-4}$\\
Case 3	& $2\pi$		& $2\pi$		& $1.00$		& $128$		& $128$		& $128$		& $0.1$	& $1.37$			& $2\times10^{-4}$\\
Case 4	& $2\pi$		& $2\pi$		& $1.00$		& $128$		& $128$		& $128$		& $0.1$	& $0.98$			& $2\times10^{-4}$\\
Case 5	& $2\pi$		& $2\pi$		& $1.00$		& $128$		& $128$		& $128$		& $0.1$	& $0.59$			& $2\times10^{-4}$\\
Case 6	& $2\pi$		& $2\pi$		& $1.00$		& $128$		& $128$		& $128$		& $0.1$	& $0.20$			& $2\times10^{-4}$\\
\hline
\end{tabular}
\end{center}
\label{Table1}
\end{table*}

\begin{table*} 
\caption{Summary of variables in this work.}
\begin{center}
\begin{tabular}{ccc}
\hline
Variable 			& Definition						& Formula					\\
\hline
$H$				& Domain height					& /						\\
$L_{x}$			& Streamwise extent 				& /						\\
$L_{y}$			& Spanwise extent 					& /						\\
$L_{z}$			& Wall-normal extent 				& /						\\
$N_x$			& Streamwise grid number			&/						\\
$N_y$			& Spanwise grid number				&/						\\
$N_z$			& Wall-normal grid number			&/						\\
$h$				& Element height 					& /						\\
$\tilde{.}$			& Grid-filtered quantity				& /						\\
$\hat{.}$			& Conditionally-averaged quantity 		& /						\\
$\langle{.}\rangle_{n}$& Averaged quantity on $n$ dimension 	& /						\\
$.^{\prime}$		& Fluctuating quantity 				& $.-\langle{.}\rangle_t$		\\
$\boldsymbol{u}(\boldsymbol{x},t)$ & Flow velocity			& /						\\
$\langle{\tilde{\omega}_z}\rangle_t$ & Reynolds-averaged wall-normal vorticity & ${\partial_x\langle{\tilde{v}}\rangle_t}-{\partial_y\langle{\tilde{u}}\rangle_t}$ \\
$P(x)$			& Probability density function			& /						\\
$Q(\boldsymbol{x})$	& Q criterion 						& $\dfrac{1}{2} \left( \boldsymbol{\mathsf{S}} \boldsymbol{:} \boldsymbol{\mathsf{S}} - \boldsymbol{\mathsf{\Omega}} \boldsymbol{:} \boldsymbol{\mathsf{\Omega}} \right)$ \\
$\boldsymbol{\mathsf{\Omega}}$		& Rotation rate tensor				& $\frac{1}{2} \left(\nabla \tilde{\boldsymbol u} - \nabla \tilde{\boldsymbol u}^\mathrm{T} \right)$ \\
$\boldsymbol{\mathsf{S}}$			& Strain rate tensor					& $\frac{1}{2} \left(\nabla \tilde{\boldsymbol u} + \nabla \tilde{\boldsymbol u}^\mathrm{T} \right)$ \\
$\lambda$			& Hairpin shedding distance			& $\lambda \sim h(\boldsymbol{x})$\\
$u_*(\boldsymbol{x},t)$ & Friction velocity 				& $(\delta_z |\boldsymbol{f}(\boldsymbol{x},t)|)^{1/2}$\\
\hline
\end{tabular}
\end{center}
\label{Table2}
\end{table*}

\section{Results}
\label{s:results}

In this section, the results and data from LES simulation will be displayed to demonstrate the turbulent flow characteristics over the series of transversal bars. Figure 2 displays spatial averaged first order statistics. The monotonic increasing body force is captured due to enhanced obstructive effects of roughness elements. From Figure 3 to Figure 6, first order statistics in vertical plane is used to demonstrate the turbulent coherence changes and momentum transfer between inner and outer layers. As streamwise spacing decreases, the inclination angle gets elevated to $15^{\circ}$, which is the value detected in channel flow \citep{awasthi2018numerical}. Meanwhile, probability density function of three fluctuating variables demonstrate the limitation of increasing roughness element numbers. On the one hand, the increasing element number will induce enhanced shear flow in the leeward; on the other hand, condensed element distribution will suppress wake recovery in wake. From \citep{wang2019turbulence}, we understand the downwind obstruction will decrease the spatial coherence of vortex shedding. Thus, Figure 8 illustrate the decreasing length scale of swirling structures within roughness sublayer. 

\begin{figure}
\begin{center}
\includegraphics[width=10cm]{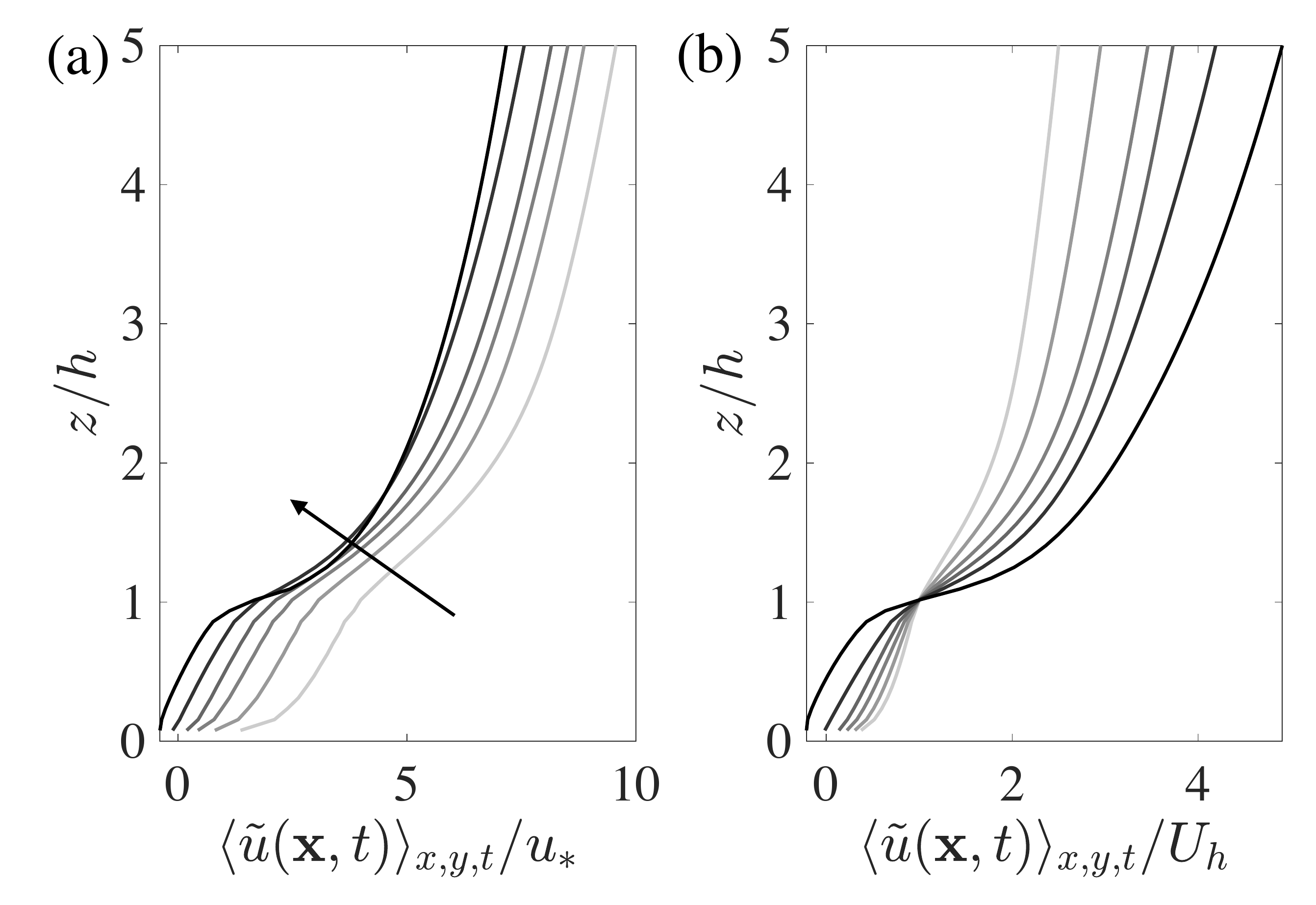}
\caption{The horizontal plane and Reynolds-averaged streamwise velocity $\langle{\tilde{u}(\boldsymbol{x},t)}\rangle_{x,y,t}$ profile against wall-normal elevation. Panel (a) shows the $\langle{\tilde{u}(\boldsymbol{x},t)}\rangle_{x,y,t}$ profile normalized by universal friction velocity. The black quiver indicates the $\lambda/H$ increasing trend. While, Panel (b) shows the $\langle{\tilde{u}(\boldsymbol{x},t)}\rangle_{x,y,t}$ normalized by the mean streamwise velocity at the canopy height $U_h$. The solid light gray to black line is Case 1 to 6 respectively.}
\label{Figure2}
\end{center}
\end{figure}

Figure 2 shows the horizontal plane and Reynolds-averaged streamwise velocity profiles against wall-normal elevation. Velocity is nondimensionalized by global friction velocity (Panel (a)) and global velocity at canopy elevation (Panel (b)), respectively. The global friction velocity $u_*\approx 0.9 m/s$ is retrieved from the maximum shear (the value at canopy height). The velocity magnitude keeps decreasing with $\lambda_x$ at all elevations. The increasing roughness element number, effectively, enhances the drag force in the whole channel. Meanwhile, we can easily capture the displacement height of roughness sublayer changes from Case 1 to 6. The log law profile is merely validate beyond the displacement height, which is also called inertial sublayer. Previously, \citep{wang2019turbulence} has revealed the vertical delamination of boundary layer over open-wide dune field. The results indicate turbulent mixing process associated with smaller length scale turbulent eddies are dominated within the roughness sublayer, reaching to $2-3$ dune field height. The inertial layer, overlaying on the roughness sublayer, keeps channel flow turbulence characteristics. While, with the monotonically intensified inhomogeneity, Case 6 shows less effectiveness. With roughness element density increasing, $\lambda_x$ decreases. The highest density is corresponded to $\lambda_x=0$, which is pure channel flow. Thus, such increasing trend of drag enhancement will not grow continuously. When $\lambda_x=2.16$ jumps to $\lambda_x=2.94$, this increment $\delta \lambda_x = 0.78$ is most effective compared with any other. While, Panel (b) shows the momentum transport pattern behind this. With $\lambda_x$ increasing, more momentum flux enveloped in roughness sublayer will get transported to higher elevations. From \citep{wang2019turbulence}, we knew the interactive process between different sublayers in boundary layer is achieved by different types of structures: from medium to large scale, swirling structures such as ``hairpin'' vortices; in small length scales, turbulent ejections are prosperous. 

\begin{figure}
\begin{center}
\includegraphics[width=16.5cm]{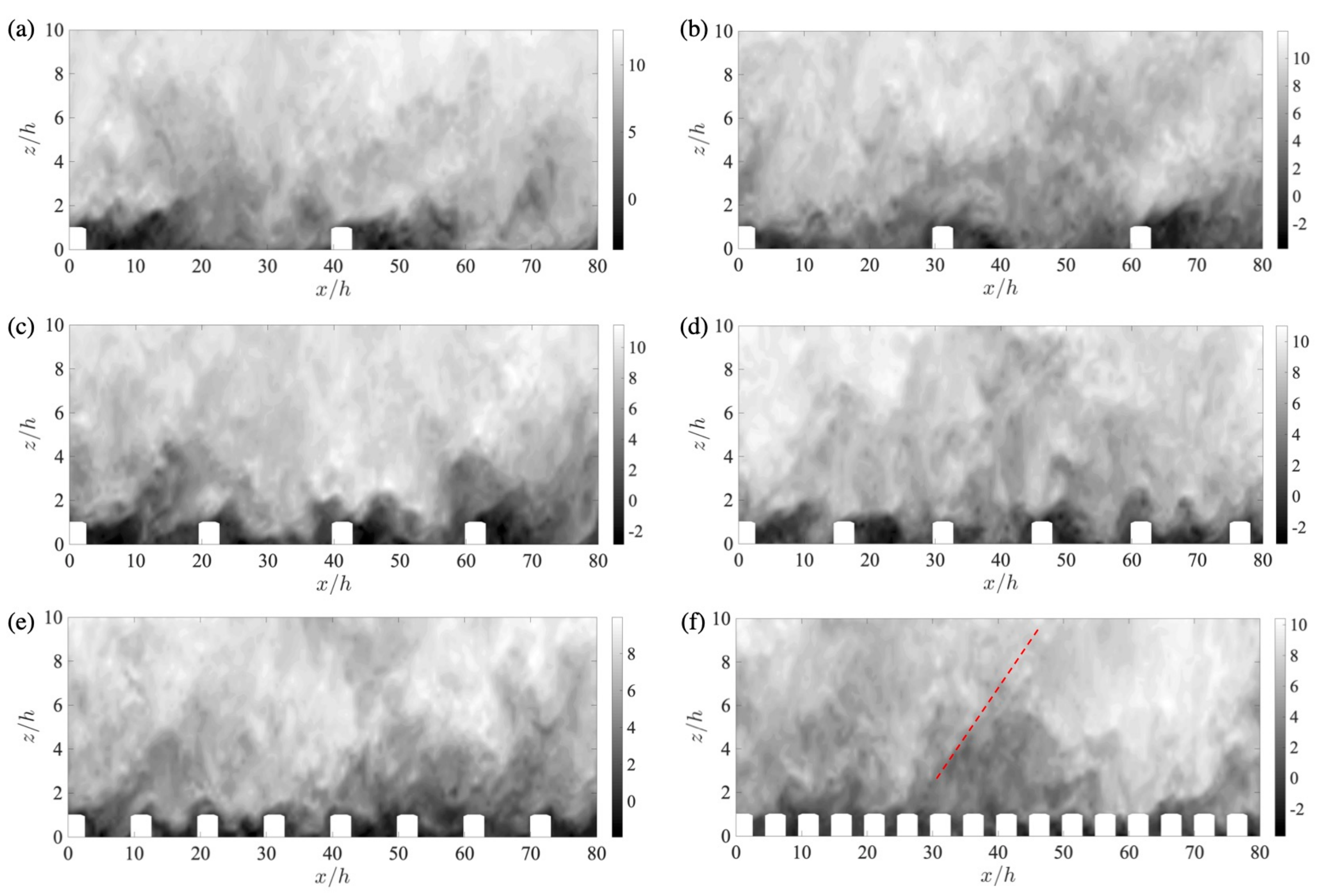}
\caption{Instantaneous streamwise velocity in the central streamwise--wall-normal plane $\tilde{u}(x,y=L_x/2,z)/u_*$. From Panel (a) to (f) are Case 1 to Case 6, respectively. Red dashed line indicate the inclination angle which is close $15^{\circ}$.}
\label{Figure3}
\end{center}
\end{figure}

Figure 3 shows the instantaneous streamwise velocity in the streamwise--wall-normal plane. Turbulent mixing process is evidently surrounding the roughness element beneath a certain height, around $5-6$ times element height. From Panel (a), we can see the typical flow pattern under low $\lambda_x$. Due to roughness element, shear layer is enveloped in a certain slope, with low velocity magnitude region in lee side. The large $\lambda_x$ provide long enough distance for turbulent flow to recover and turn to channel flow regime. When increasing element number, it will be harder for leeward turbulent flow to recover its length scale and velocity profile. Thus, in Figure 6, shorter recirculating bubble behind element with $\lambda_x$ increasing indicate the lower recovering ratio in highly condensed roughness pattern. In Panel (e) and Panel (f), the low momentum region is packed around the roughness not only in the leeward region. \citep{wang2019turbulence} has also revealed the turbulence structure in roughness sublayer. In recovering region, separated flow grows with increasing turbulent length scale and velocity magnitude. From two-point correlation map in \citep{wang2019turbulence}, the integral length profiles are retrieved and shows mixing-layer-type of attributes within roughness sublayer, where increasing and decreasing pattern has been found in the separation zone, but channel flow attributes (attached-eddy hypothesis) in inertial sublayer. Meanwhile, in Figure 3, as $\lambda_x$ decreases, packed and mixed turbulent eddies emerge as large length scale, which is consistent with Figure 1 Panel (b). Condensed roughness topography agitates enhanced vertical momentum transfer. Via incremental shear production within roughness sublayer, large length scale eddy can grow and extent to lower elevations close to canopy height, which keeps feeding smaller length scale eddies beneath them. The inclination angle in Panel (a) is less than $15^{\circ}$. However, with element number increasing, channel flow characteristics get recovered, the inclination angle goes to $15^{\circ}$ \citep{salesky2018buoyancy}.

\begin{figure}
\begin{center}
\includegraphics[width=16.5cm]{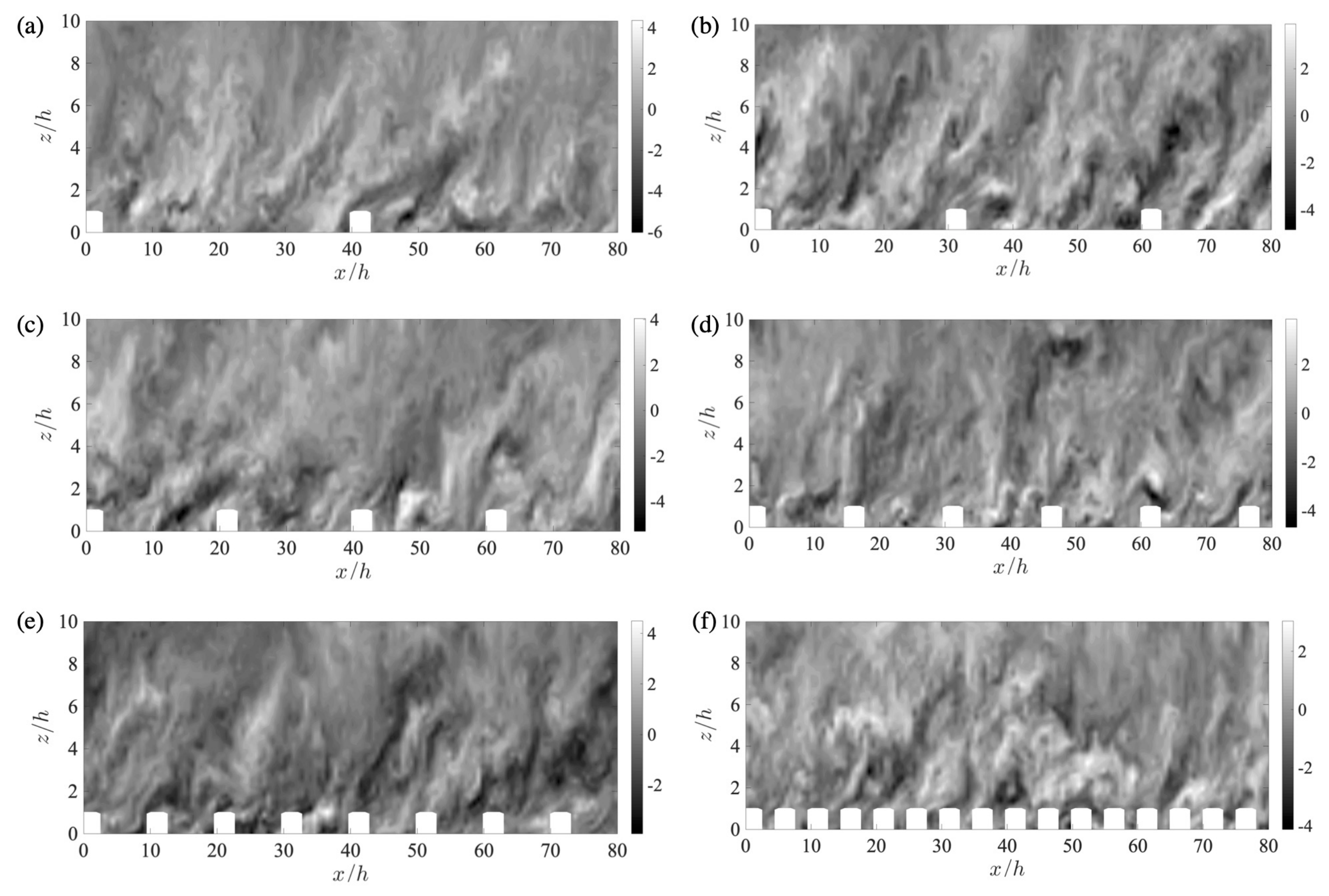}
\caption{Instantaneous spanwise velocity in the central streamwise--wall-normal plane $\tilde{v}(x,y=L_x/2,z)/u_*$. From Panel (a) to (f) are Case 1 to Case 6, respectively.}
\label{Figure4}
\end{center}
\end{figure}

Figure 4 shows the instantaneous spanwise velocity in the central streamwise--wall-normal plane. It shows the evidence of spanwise turbulent mixing, although spanwise homogeneity is achieved in topography. The spanwise turbulent mixing traces can extend to $6-8$ roughness element height. Inclination angle is clearly captured, which still displays a increasing trend as $\lambda_x$ decreases. In low compacting topography cases ($\lambda_x/h>1.00$), spanwise turbulent footprints are evidently under the topography compacting ratio effects. However, when $\lambda_x/h<1.00$, spanwise mixing starts to analogy channel flow-type of spanwise instability, that is consistent with Figure 3. 

\begin{figure}
\begin{center}
\includegraphics[width=16.5cm]{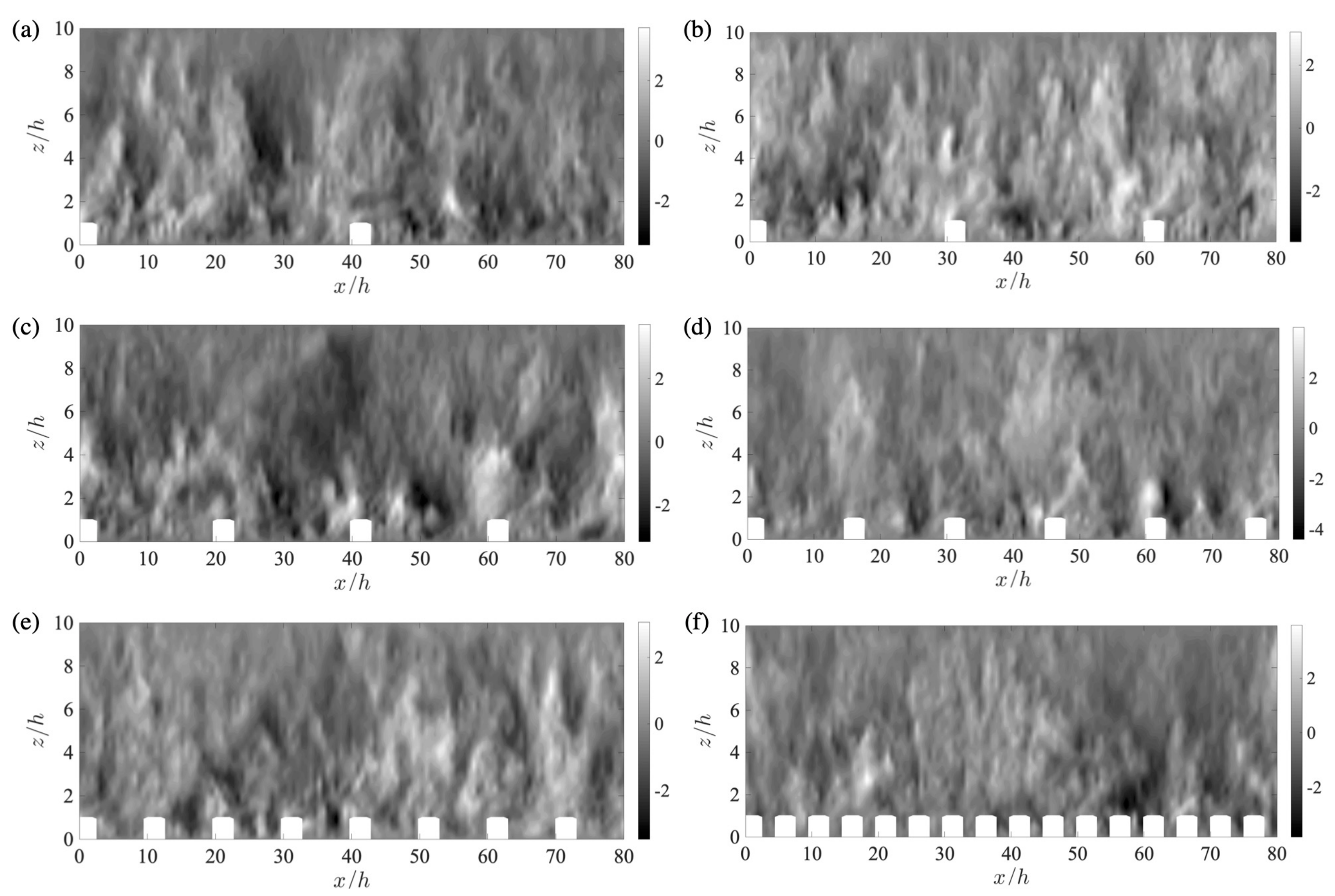}
\caption{Instantaneous wall-normal velocity in the central streamwise--wall-normal plane $\tilde{w}(x,y=L_x/2,z)/u_*$. From Panel (a) to (f) are Case 1 to Case 6, respectively.}
\label{Figure5}
\end{center}
\end{figure}

Figure 5 displays instantaneous wall-normal velocity in the central streamwise--wall-normal plane. The magnitude of wall-normal velocity is approximately same with spanwise velocity, which indicates same turbulent momentum transport strength in spanwise and wall-normal direction. The mixing process gets enhanced as displacement height increases if $\lambda_x/h<1.0$. For $\lambda_x/>1.0$, such trend gets inverted, where more compacting topography will trigger lower turbulent mixing progress and will analogy low roughness compacting ratio cases. Such instantaneous visualization is consistent with the Reynolds-averaged results (Figure 6) and probability density function (PDF) results (Figure 7). 

\begin{figure}
\begin{center}
\includegraphics[width=16.5cm]{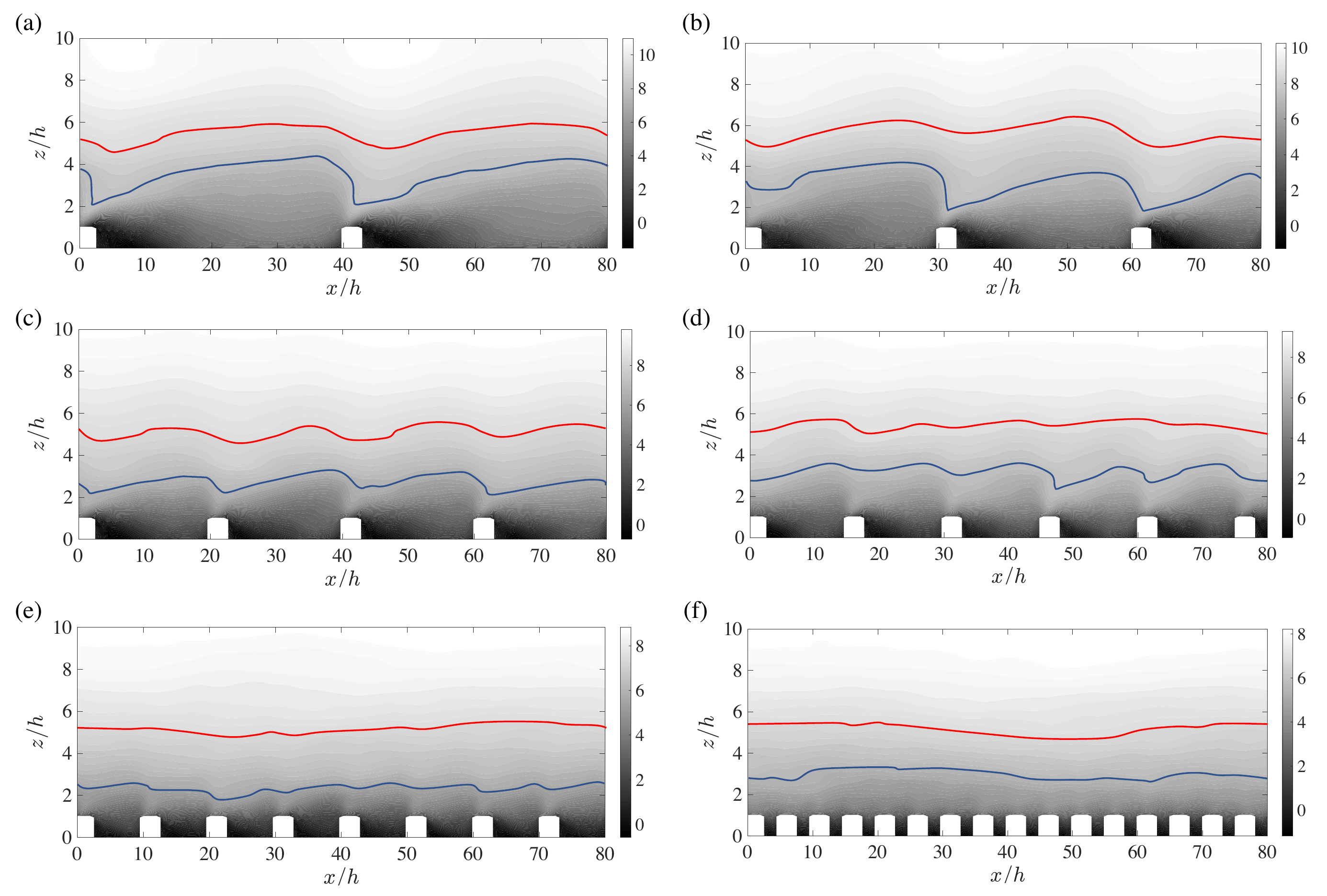}
\caption{Reynolds-averaged streamwise velocity in the central streamwise--wall-normal plane $\langle \tilde{u}(x,y=L_x/2,z)\rangle_{t}/u_*$. From Panel (a) to (f) are Case 1 to Case 6, respectively. Solid red lines indicate velocity isoline at $z/h\approx5.8$. While, solid blue lines indicate velocity isoline at $z/h\approx2.8$.}
\label{Figure6}
\end{center}
\end{figure}

Figure 6 shows Reynolds-averaged streamwise velocity in the central streamwise--wall-normal plane, which demonstrates the monotonically heterogeneity at turbulent shear layer. Panel (a) is Reynolds-averaged streamwise velocity in Case 1. It shows a ``curved'' flow feature at element leeward, which is the signature of shear layer. This shear ``curve'' can extent to whole domain region ($8-10h$) in Case 1. And the shear ``curve'' in Case 2 (Panel (b)) and Case 3 (Panel (c)) can all extent to $10h$. While, shear ``curve'' effective region gets lower in Case 4 (Panel (d)). A decreasing trend of shear ``curve'' region height is evident in Case 5 and 6 (Panel (e,f)). In Panel (e), the effective region can merely extent to $3h$. While, Panel (f) shows no shear effective region under Reynolds-averaged results. In \citep{wang2019turbulence}, roughness sublayer region has been defined  as $3-5$ times element heights. While, the dune field roughness sublayer can range from $2-3$ times dune field height \citep{wang2019thesis}. We can see element compacting ratio is a very significant factor to determine the roughness sublayer region and shear layer strength via Figure 6. Remember that \citep{leonardi2004structure} reports as $w/k$ increases, the coherence decreases in the streamwise direction having a minimum for $w/k=7$. That is because most outward motion occurring close to the leading edge of the elements. Meanwhile, the Reynolds stress anisotropy tensor and its invariants show a closer approach to isotropy over the rough wall than over a smooth wall. The solid red lines indicate the flow pattern at outer region. The plot at $z/h\approx5.8$ demonstrates the outer region is transferring from mixing layer to outer layer. The displacement height keeps decreasing in Panel (e) and Panel (f). The isoline at $z/h\approx2.8$ shows more obvious transferring trend. 

From these aforementioned results, the evident effects of roughness element distribution is demonstrated here. To isolate three dimensional instability, spanwise homogeneity has been achieved via the transversal features. Thus, it is hard to capture the transversal swirling motions such as secondary flow. Based on the plane average of first order statistics, roughness element drag effect is increased as element distribution becomes condenser. However, such increasing trend is impaired by continuous decreasing of $\lambda_x$. The consistent attribute is revealed from instantaneous flow visualization too. In Figure 3, 4 and 5, large eddy breaks down into smaller aggregations of smaller eddies with results of lower turbulent coherences within roughness sublayer. Those ``enhancing-wakening'' trend is a signature of the evidence that canopy flow is transferring to channel flow. In canopy flow such as the turbulent flow over buildings, dunes, vegetations, turbulent flow within roughness sublayer displays an evident attribute that is mixing layer analogy \citep{wang2019turbulence}, where the spatial coherence length scale is proportional to mixing layer length scale. While, mixing layer eddies can spatially grow with $\lambda_x$ decreasing and transfer into attached eddy, which displays wall-normal elevation correlation. \citep{yang2018numerical} also recorded such transitioning progress in streamwise homogeneous topographies, wherein, under high roughness element distribution cases, turbulence statistics display characteristics very similar to a homogeneous roughness and thus the surface may be regarded as a roughness. While, under sparse conditions, secondary flow patterns shows highly correlated to roughness element spacings. It can be verified via the evidence showed in Figure 6.

\begin{figure}
\begin{center}
\includegraphics[width=16.5cm]{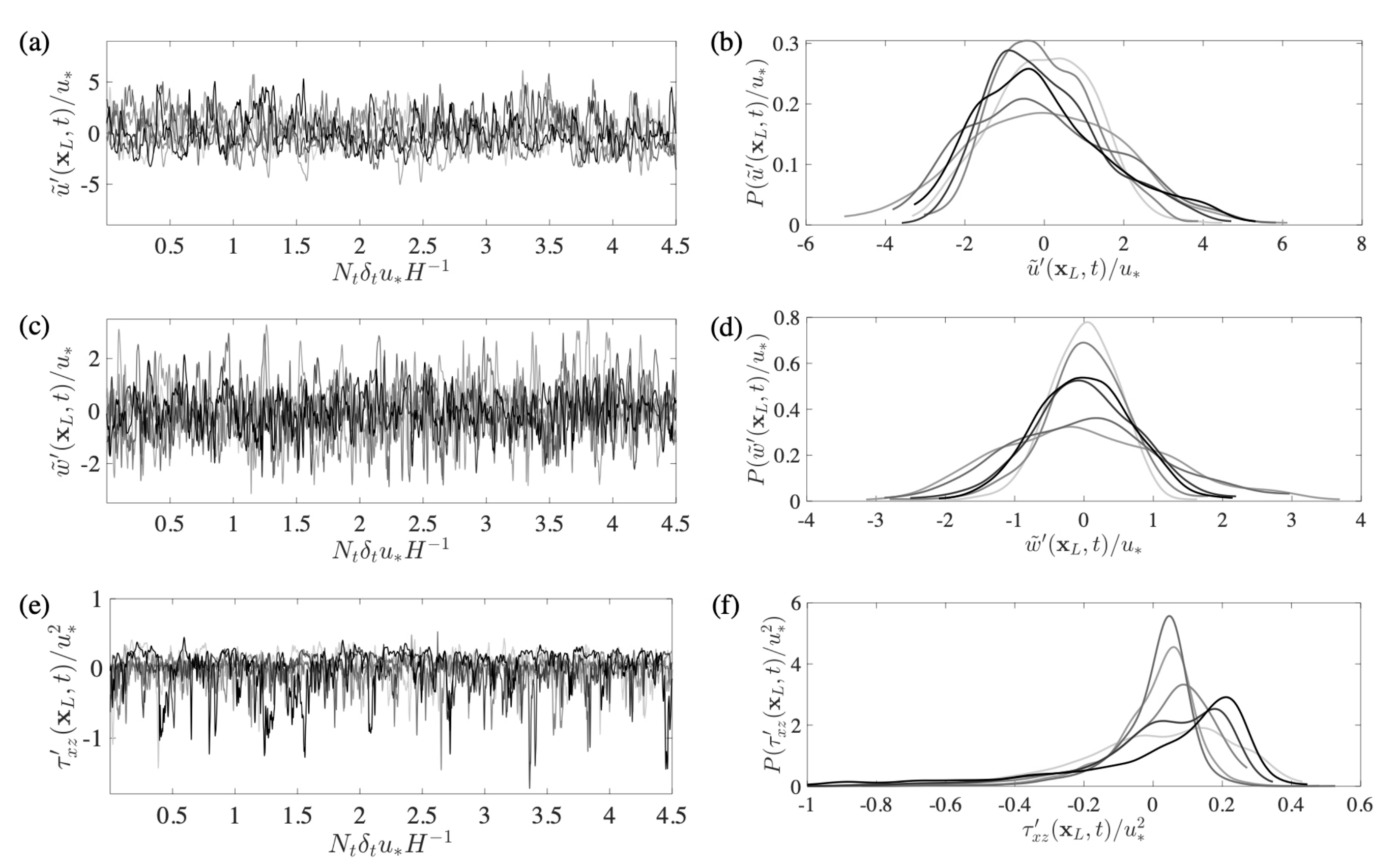}
\caption{Panel (a,c,e) are time-series of streamwise fluctuating velocity $\tilde{u}^{\prime}(\boldsymbol{x}_L,t)/u_*$, wall-normal fluctuating velocity $\tilde{w}^{\prime}(\boldsymbol{x}_L,t)/u_*$, and streamwise--wall-normal wall stress fluctuation ${\tau}_{x,z}^{\prime}(\boldsymbol{x}_L,t)/u_*^2$ on the sampling point $\boldsymbol{x}_L=\{x=L_x/2,y=L_y/2,z=h\}$. Panel (b,d,f) are PDF of them respectively. The solid light gray line to black line is Case 1 to 6 respectively.}
\label{Figure7}
\end{center}
\end{figure}

Figure 7 shows the time-series of streamwise fluctuating velocity $\tilde{u}^{\prime}(\boldsymbol{x}_L,t)/u_*$, wall-normal fluctuating velocity $\tilde{w}^{\prime}(\boldsymbol{x}_L,t)/u_*$, and streamwise--wall-normal wall stress fluctuation ${\tau}_{x,z}^{\prime}(\boldsymbol{x}_L,t)/u_*^2$ on the sampling point $\boldsymbol{x}_L=\{x=L_x/2,y=L_y/2,z=h\}$ in Panel (a), (c) and (e), respectively. While, Panel (b), (d) and (f) are probability density function (PDF) of three variables in Panel (a), (c) and (e), respectively. From the solid light gray to black is from Case 1 to Case 6, respectively. The magnitude of normalized streamwise velocity fluctuation at Point $\boldsymbol{x}_L$ can approach $5.0$. While, the non-dimensionalized wall-normal velocity fluctuation magnitude is more than $2.0$. In Panel (a) and (c), the magnitude varies in different cases. Consistently, maximal values emerges in Case 2 and 3 in Panel (a) and (c). It is more evident in corresponding PDF profiles (Panel (b,d)), where, as $\lambda_x$ decreases, PDF profile gets wide with its peak lower from Case 1 to Case 3. However, when $\lambda_x/h\approx1.0$, PDF profile gets narrow and peak increases again, with $\lambda_x$ continuous increasing. Due to cases number limitation, if $\lambda_x$ keeps decreases after $\lambda_x/h=0.2$, the light gray line could collapse black line well. From PDF profiles, we can easily capture the maximum magnitude value of streamwise and wall-normal velocity fluctuation is $6.0$ and $3.7$, respectively, which is all appears in Case 2. Thus, the effectiveness of spacing changing is extraordinary from Case 1 to Case 2. But after that the effective range is weakened, because, on the one hand, less $\delta \lambda_x$ will limit such tendency, on the other hand, homogeneous roughness attributes emerges after Case 2 \citep{yang2018numerical}. All PDF profiles have none skewness. 

Figure 7 Panel (e) and (f) display the PDF of stress fluctuation,
\begin{equation*}
{\tau}_{x,z}^{\prime}(\boldsymbol{x}_L,t)/u_*^2={\tau}_{x,z}(\boldsymbol{x}_L,t)-\langle{{\tau}_{x,z}(\boldsymbol{x}_L,t)}\rangle_{t}=\nu_t\left(\frac{\partial\tilde{w}(\boldsymbol{x}_L,t)}{\partial x}-\frac{\partial\tilde{u}(\boldsymbol{x}_L,t)}{\partial z}\right)-\nu_t\left\langle{\left(\frac{\partial\tilde{w}(\boldsymbol{x}_L,t)}{\partial x}-\frac{\partial\tilde{u}(\boldsymbol{x}_L,t)}{\partial z}\right)}\right\rangle_t.
\end{equation*}
From the time series of ${\tau}_{x,z}^{\prime}(\boldsymbol{x}_L,t)/u_*^2$, we can see the maximum magnitude does not exists in Case 2 or 3, instead in Case 1 and 6. PDF profile of ${\tau}_{x,z}^{\prime}(\boldsymbol{x}_L,t)/u_*^2$ shows an opposite trend to Panel (b) and (d). In Panel (f), as the $\lambda_x$ decreases, PDF profile becomes narrow, which indicates that less high magnitude fluctuation exists on Point $\boldsymbol{x}_L$. However, the large stress fluctuations come up again after Case 3. This ``decreasing-increasing'' stress fluctuation patten from Case 1 to 6 demonstrates that high stress fluctuations will emerge more frequently in channel flow. Note, here, the skewness of PDF profiles also present a similar ``decreasing-increasing" tendency. From Case 2 to 6, the skewness of PDF profile increases. Meanwhile, Case 1 also shows positive skewness. Those indicate that in channel flow, inner and outer interaction ($Q1$ and $Q3$) are dominant, not like in canopy flow where $Q2$ and $Q4$ are dominant \citep{AndersonChamecki14,wang2019thesis}.

\begin{figure}
\begin{center}
\includegraphics[width=16.5cm]{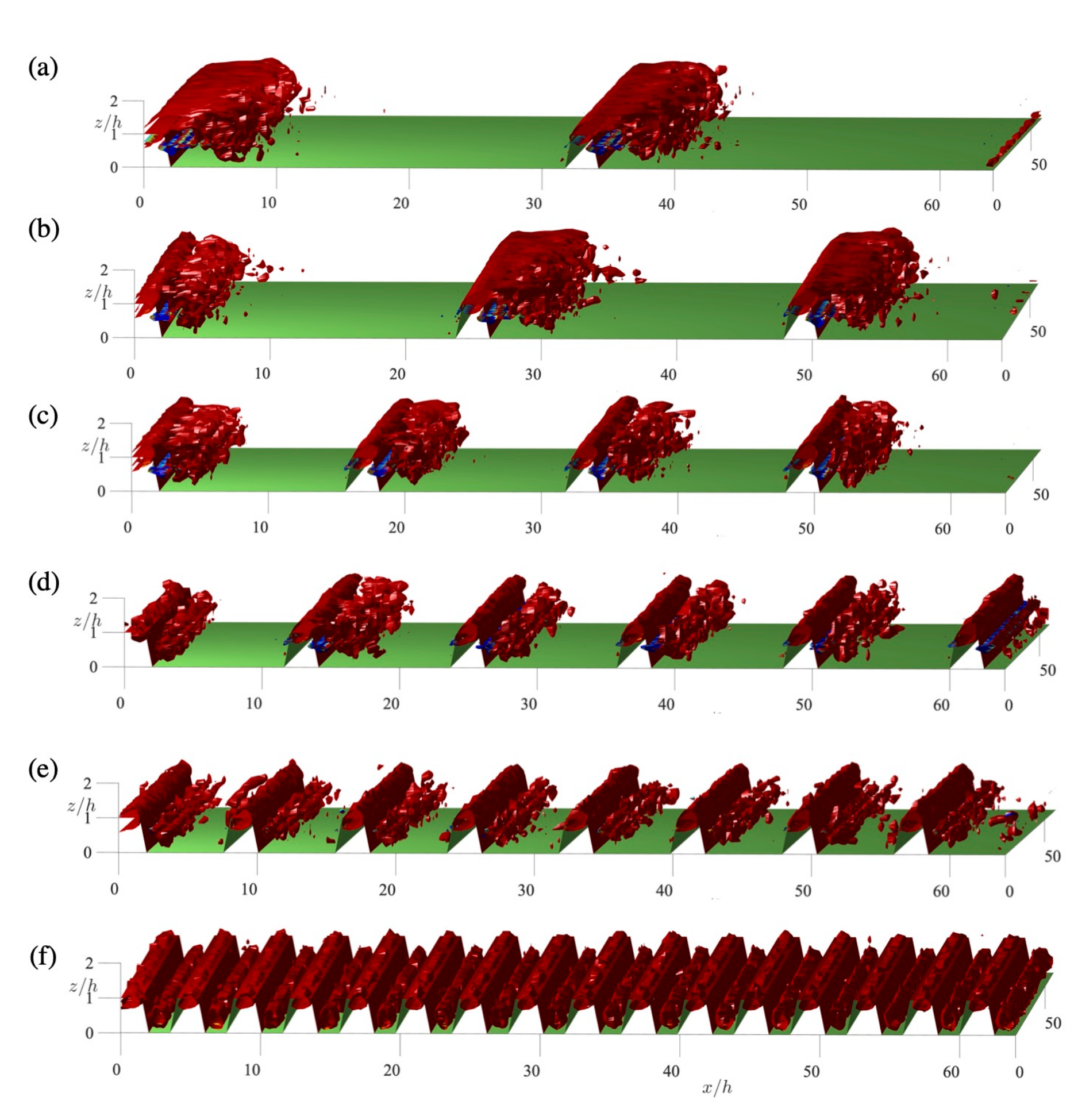}
\caption{The Reynolds-average $Q$ criterion isosurfaces in Case 1 to 6, respectively. The color indicates the spanwise rotating direction. Red indicates positive spanwise rotating. While, blue indicates negative spanwise rotating. From Panel (a) to (f) are Case 1 to 6, respectively.}
\label{Figure8}
\end{center}
\end{figure}

From Figure 3 to Figure 7, 1D and 2D data have been exhibited to demonstrate the transferring tendency during $\lambda_x$ changes. In Figure 8, three dimensional flow visualization are presented and we compute the $Q$ criterion vortex identifier, which is derived from the velocity gradient tensor, $\boldsymbol{\mathsf{D}} = \nabla \tilde{\boldsymbol u}$ \citep{jeonghussain,christensenadrian01}. $\boldsymbol{\mathsf{D}}$ can be decomposed into its symmetric and anti-symmetric components, $\boldsymbol{\mathsf{D}} = \boldsymbol{\mathsf{S}} + \boldsymbol{\mathsf{\Omega}}$, where $\boldsymbol{\mathsf{S}} = \frac{1}{2} \left(\nabla \tilde{\boldsymbol u} + \nabla \tilde{\boldsymbol u}^\mathrm{T} \right)$ and $\boldsymbol{\mathsf{\Omega}} = \frac{1}{2} \left(\nabla \tilde{\boldsymbol u} - \nabla \tilde{\boldsymbol u}^\mathrm{T} \right)$, allowing computation of the $Q$ criterion with:
\begin{equation*}\label{Qequ}
Q = \dfrac{1}{2} \left( \boldsymbol{\mathsf{S}} \boldsymbol{:} \boldsymbol{\mathsf{S}} - \boldsymbol{\mathsf{\Omega}} \boldsymbol{:} \boldsymbol{\mathsf{\Omega}} \right).
\end{equation*}
In Figure 8, red color indicates the positive spanwise rotation, while blue indicates negative spanwise rotation. The isosurfaces reveal the swirling coherent structures. From Panel (a) to (f), they correspond to Case 1 to Case 6, respectively. In Figure 8, we can see, firstly, the coherent structure length scale keeps decreasing from Case 1 to Case 6, which is consistent with Figure 6. Due to decreased successive spacing, large coherent eddies breaks into aggregations of smaller length scale eddies \citep{wang2019turbulence}. Secondly, the effective region of mixing layer keeps shrinking with $\lambda_x$ decreasing. In Case 1, the large coherent eddy over the element can range to $2h$. However, in Case 6, it is less than $1.2h$. Meanwhile, we can see the flow structure is highly correlated to element spacing from Case 1 to 3. Finally, in Case 1 to Case 4, we can capture the spanwise counterrotating in leeward wake region. However, after Case 5, flow pattern turns into a series of successive spanwise rollers, the diameter of which is approximately close to $0.5h$. Thus, Figure 8 verify these aforementioned results and decently reveals the actual flow swirling structures with roughness sublayer. 

\section{Conclusion}
\label{s:conclusion}
Kevin-Helmholtz instability triggered hairpin vortex shedding has been widely concluded as the ``signature'' of mixing layer analogy in most of canopy flow regimes. However, convoluted roughness types complicate the observing practice of turbulent evolving progress. To simplify that, spanwise homogeneous roughness has been adopted in this work to capture the streamwise flow alternations under different streamwise roughness element distances. Streamwise successive object distance $\lambda_x$ is used to show the roughness element distance. With the element number increasing, due to element obstructive effects, drag force increases with $\lambda_x$ decreasing via the observation of first order statistic analysis. Case 1 shows heterogeneous flow pattern happens under sparse element distributions. But Case 6 shows homogeneous flow pattern. With successive spacing decreasing, the transitional tendency from turbulent channel flow to canopy flow has been observed from Case 1 to Case 4. However, when $\lambda_x/h<1.0$, as successive distance decreases, the evidences indicate turbulent flow is transferring from canopy flow into channel flow from Case 4 to Case 6. Instantaneous results verifies the enhancement of turbulent mixing and decreasing turbulent coherency. The attached-eddy hypothesis (AEH) becomes valid in $\lambda_x/h<1.0$ Cases. Meanwhile, Reynolds-averaged results show the roughness shear ``curve'' effectiveness will decrease when $\lambda_x$ keeps decreasing after $\lambda_x/h<1.0$. The probability density function (PDF) of three fluctuating variables on a sampling point over element display the ``increasing-decreasing'' trend of decreasing spacing effect on upper region, wherein the maximal effect emerges in Case 2 and 3. And the swirling structural visualizations also demonstrate the progress of turbulent structure changing in roughness sublayer at different streamwise successive object distances. All results comprehensively show a great consistent evidence, that is the channel flow-analogy emerges dense roughness canopy when $\lambda_x<1.0$, where topographies show roughness effects. Accordingly, $\lambda_x$ is important to decide the flow features including displacement height, surface shear strength, turbulent structures and so forth. However, due to the limited case number, $\lambda_x=1.0$ could be a crude dividing value to decide the flow regime. 

\bibliography{mybib.bib}  

\end{document}